\documentclass[sigconf]{acmart}

\AtBeginDocument{%
  \providecommand\BibTeX{{%
    \normalfont B\kern-0.5em{\scshape i\kern-0.25em b}\kern-0.8em\TeX}}}

\copyrightyear{2022} 
\acmYear{2022} 
\setcopyright{rightsretained} 
\acmConference[CUI 2022]{4th Conference on Conversational User
Interfaces}{July 26--28, 2022}{Glasgow, United Kingdom}
\acmBooktitle{4th Conference on Conversational User Interfaces (CUI
2022), July 26--28, 2022, Glasgow, United
Kingdom}\acmDOI{10.1145/3543829.3544531}
\acmISBN{978-1-4503-9739-1/22/07}
\begin{document}

\title[Embrace your incompetence!]{Embrace your incompetence! Designing appropriate CUI communication through an ecological approach}

 
\author{Sophie Becker}
\affiliation{%
  \institution{University College Dublin}
  \city{Dublin}
  \country{Ireland}}
\email{sophie.becker@ucdconnect.ie}
 
\author{Philip R. Doyle}
\affiliation{%
  \institution{University College Dublin}
  \city{Dublin}
  \country{Ireland}}\email{philip.doyle1@ucdconnect.ie}

\author{Justin Edwards}
\affiliation{%
  \institution{University College Dublin}
  \city{Dublin}
  \country{Ireland}}\email{justin.edwards@ucdconnect.ie}


\begin{abstract}
 People form impressions of their dialogue partners, be they other people or machines, based on cues drawn from their communicative style. Recent work has suggested that the gulf between people's expectations and the reality of CUI interaction widens when these impressions are misaligned with the actual capabilities of conversational user interfaces (CUIs). This has led some to rally against a perceived overriding concern for naturalness, calling instead for more representative, or appropriate communicative cues. Indeed, some have argued for a move away from naturalness as a goal for CUI design and communication. We contend that naturalness need not be abandoned, if we instead aim for ecologically grounded design. We also suggest a way this might be achieved and call on CUI designers to embrace incompetence! By letting CUIs express uncertainty and embarrassment through ecologically valid and appropriate cues that are ubiquitous in human communication - CUI designers can achieve more appropriate communication without turning away from naturalness entirely. 
\end{abstract}



\maketitle
\textbf{}
\section{Introduction}
Conversational User Interface (CUI) research has readily identified a sense of weirdness and artificiality about socio-cognitive aspects of human-machine dialogue (HMD) interaction \cite{cowan2017can, meah2014uncanny, doyle_mapping_2019}, along with a general dissatisfaction with usability and the user experience \cite{luger_like_2016,leahu2013categories,ammari2019music,doyle_mapping_2019}. Indeed, the acuteness of dissatisfaction some users experience is often said to result from a gulf between their expectations for the interaction and the realities of the way current widely adopted CUIs function and behave \cite{luger_like_2016,clark_what_2019}. Closing the gap between expectation and reality would significantly improve CUI interactions for millions of people who currently use them, but this is not an easy task. Recently a growing body of work has advocated for increased focus on appropriateness in CUI design. Predominantly this revolves around the notion of designing systems with appropriate levels of human-likeness, though fundamentally the idea is that a system’s features such as name, voice and communicative style should be congruent with its actual capabilities \cite{moore_appropriate_2017}. That is, to design systems that behave in a way that might be expected from an advanced machine that can talk, rather than mimicking dialogue behaviours seen in human-human dialogue (HHD) \cite{aylett2019right,aylett2019siri}. This is intended to shift the focus from designing styles of communication that seem natural - a sometimes myopic goal, particularly in synthesis development -  toward designing appropriate ones \cite{aylett2019right,aylett2019siri,lem2021easy}. Essentially, this places appropriateness and naturalness at odds; competing design goals aimed at shaping the path to improved CUI user experiences. 

Here, we build on these ideas and propose an alternative approach: ecological design, an approach that leverages the most natural and intuitive aspects of the way people speak in HHD, whilst always remaining cognizant about the nature of the dynamic (i.e., humans and machines working collaboratively through dialogue) and the expectations that stem from this. One way this might be achieved is by developing systems that are designed to embrace their incompetence. We argue that errors in communication are perfectly natural between humans and need not be treated as a phenomenon unique to dialogue with machines. Instead, we contend that machines can and should signal their capacity for error just like humans do, pursuing appropriateness not by abandoning naturalness, but by embracing a type of context appropriate naturalness: an ecologically grounded design approach. 

\section{Partner Models and Audience Design in Human-Machine Dialogue}
Research into human conversation suggests that people adapt their speech and language choices in dialogue based on the perceived communicative competence and social relevance of their dialogue partner \cite{brennan_chapter_2010, branigan_role_2011,clark1996using,amalberti_user_1993}. Essentially, it is argued that people make informal hypotheses  about their partner’s communicative abilities and tendencies based on stereotypical assumptions drawn from superficial cues (i.e., age, ethnicity, gender, etc.) \cite{brennan_chapter_2010, branigan_role_2011}. Then, following exposure to a given partner, they develop a more nuanced heuristic or mental model of their interlocutor’s capabilities \cite{brennan_chapter_2010}. In psycholinguistics these are referred to as partner models, which are said to support successful communication by guiding a speaker toward appropriate language choices for a given partner \cite{brennan_chapter_2010, branigan_role_2011}. As such, partner modelling is closely associated with concepts like perspective taking and audience design \cite{bell1984language}, whereby speakers adopt an allocentric and empathetic stance, tailoring their speech for a specific audience (person or group).

Partner modelling effects on language production have been shown in both HHD and HMD. For the most part, partner modelling work in HMD has revolved around the impact of partner models on the phenomenon of alignment, whereby people converge on the same speech and language behaviours in dialogue. Here it has been shown that when people perceive their partner to be a ‘poor’ or ‘at risk’ \cite{branigan_role_2011,pearson2006adaptive,oviatt_linguistic_1998} dialogue partner – such as when speaking to a machine compared to a human \cite{branigan_role_2011}, or when speaking to a less advanced systems \cite{pearson2006adaptive} – they tend to align more strongly at prosodic \cite{bell2011staging,oviatt2004toward} and lexical levels \cite{branigan_role_2011, bergmann2015exploring}. Previous work has also highlighted the importance of perceived knowledge states \cite{cowan_they_2017} and intelligence \cite{luger_like_2016,amalberti_user_1993} in HMD when forming impressions about what systems are and what they can do. More recent work has looked to identify the underlying dimensions of partner models in HMD. This work has highlighted the importance of perceived competence and dependability, perceived human-likeness and perceived flexibility in communication \cite{doyle_what_2021} when evaluating systems as a distinct type of dialogue partner. Collectively the work shows that CUI design features such as a human-like voice \cite{cowan_voice_2015,moore_appropriate_2017}, word choices \cite{cowan2019s}, behaviours and dialogue strategies \cite{bell1984language,amalberti_user_1993} can impact partner models for speech interfaces and subsequent interaction behaviour. 

Indeed, interacting with a speech interface intentionally designed to seem human-like leads users to develop a different partner model than when interacting with a device designed to seem machine-like. Further, this has non-trivial implications for successful CUI interaction, particularly when the mismatch between component parts of the system is particularly large. Specifically, CUIs with human-like features such as high quality human-sounding voices and human-like names tend to lead people to believe a system is more communicatively capable than current systems prove to be \cite{cowan_voice_2015,doyle_what_2021}. Initially this might lead people to use language intuitively, as if speaking to another person, however, this is quickly revealed as being overly complex and inappropriate for the system \cite{luger_like_2016,moore2017spoken}. This mechanism - that people form partner models and use them to choose their strategies for interactions with machines - is crucial in understanding why setting proper expectations matters and, in a general sense, how doing so might improve CUI interactions.

\section{Appropriate Expectations}
Conversational user interfaces, particularly those that use speech, are generally poor at supporting social conversation, and cannot keep up with the complexity, interactivity or dexterity of free-flowing human speech \cite{clark_what_2019,doyle_mapping_2019}. Instead, they are better equipped for simple, reactive, question-answer interactions \cite{clark_what_2019,doyle_mapping_2019,cowan_they_2017,porcheron_voice_2018}. Yet, voice assistants are consistently and overtly designed, and indeed marketed, to imply a heightened sense of intelligence and social relevance, largely by mimicking and/or simulating human-likeness. Systems evaluations have also been largely driven by notions of ‘naturalness’ \cite{aylett2019right}; in no small part because MOS-X, a questionnaire for measuring perceived naturalness of speech synthesis, is one of the few relatively widely used tools of this nature in HMD research \cite{polkosky2003expanding}. In a sense, the concern with naturalness in speech synthesis design has bled into discussions about conversational design. Because humans perceive other humans to be more conversationally competent than machines, a CUI that emphasises human-likeness by design may unintentionally lead a user to believe the system is capable of advanced conversation; or at least behave as if the system is \cite{nass_computers_1994,moore1992experiences}. Here, an inflated perception of human-likeness, stemming from features other than conversational competence, might contribute to the user developing an inaccurate partner model for that system. Indeed, designing systems with humanlike characteristics that are incongruous with system capability could be described as a design fault that provides users with a poor conceptual model of system capabilities from which to develop their own mental model, undermining the chances of communicative success also. 

This creates what Norman described as the gulf between evaluation and execution \cite{norman2013design}. Not only do misaligned partner models create this gulf, but the magnitude of misalignment defines just how wide the gap is \cite{luger_like_2016,moore_appropriate_2017}. Some suggest the difference between the capabilities of current systems and the polished and professional-sounding human-like voices they use is so stark that it causes critical communication breakdowns leading to very limited use and even abandonment \cite{doyle_mapping_2019,luger_like_2016,moore_appropriate_2017}. In terms of updating one’s partner model, this encourages people to draw more polemic evaluations; moving from an impression of a system as a nuanced sophisticated communicator with idiosyncratic communicative tendencies, to being viewed as little more than a glorified paperweight. The mismatch is also known to create dissonance for users \cite{luger_like_2016,leahu2013categories}, which is cognitively burdensome and creates a frustrating user experience. Ultimately if an interlocutor has the capabilities of a computer but sounds like a human, a user will enter the interaction treating the device like it is equipped to understand human conversation only to experience disappointment when it cannot do what they were primed to expect, and helps to explain the abandonment of CUIs even by those who initially felt eager to use them \cite{luger_like_2016}. 

To address this gulf, recent research has suggested the concept of voice appropriateness: that the style of communication used by a CUI should match the functional capabilities of a given interface \cite{moore_appropriate_2017,aylett2019right,aylett2019siri,lem2021easy} . Although the design of voice agents like Google Duplex \cite{noauthor_google_2018} have continued to pursue increasingly natural and competent-sounding human-like voices, the concept of voice appropriateness interrogates whether this is the best decision for facilitating successful user interactions. Here it is suggested that by using a voice that is congruent with communicative competence  the voice can be used to manage expectations by correctly signalling system functionality to the user \cite{branigan_role_2011,cowan_they_2017,moore_appropriate_2017,moore2017spoken, aylett2019siri}. Similar enquiries in the field of robotics have demonstrated that robots designed to accurately portray their behavioural capabilities are better received by people than robots which appear to have more advanced human-like capabilities than they actually possess \cite{moore2015talking}. Recent work has argued that CUI voices might instead pursue “the right kind of unnatural” voice \cite{aylett2019right}, that they should sound robotic yet understandable in order to indicate genuine system functionality \cite{moore_appropriate_2017}. Likewise, recent work has suggested that non-human-like synthesized voices can help to achieve this goal \cite{lem2021easy}. Approaches like these, which move away from earlier concerns with ‘naturalness’ seek to foster the development of a more accurate partner model from the initial interaction, avoiding mismatched expectations and disappointment. 

A move away from naturalness has proved to be a somewhat provocative suggestion, in part because it seems to be fundamentally at odds with people’s preferences. Early work on HMD communication preferences by Nass and Lee observed evidence of a similarity-attraction effect, whereby people prefer interacting with speech systems that sound like themselves rather than like someone else or like a robot \cite{nass_does_2001}. Designing for human-likeness can also facilitate comfort and rapport building in situations where forming a bond and building trust are important, such as in healthcare, education, or social programs \cite{clark_state_2019}, and can act as a strong initial enticement for people to adopt speech interfaces in the first place \cite{luger_like_2016}. Likewise, as previously discussed, degree of human-likeness is also a key factor in how people form their partner models of dialogue partners in HMD \cite{doyle_what_2021}. 

While a growing body of research has revealed that human-like language can create feelings of distrust during a CUI interaction \cite{clark_state_2019, clark_what_2019,doyle_mapping_2019,moore2017spoken}, much of this work fails to disentangle the communication style from the communication aims. People use CUIs in order to complete a specific objective - for low-stakes tasks, for retrieving factual information, and for making multi-tasking or hands-free interactions easier \cite{ammari2019music,dubiel_survey_2018,cowan2017can,luger_like_2016}. Because people view their interactions with these systems in functional terms, they are generally not interested in building a two-way relationship with them \cite{clark_using_2002}. Rather, the conversational agent is seen as a tool for accomplishing their task, and social interaction with the device is not their primary aim \cite{clark_state_2019}. In this way, small talk, humor, and mimicry of human-likeness that fails to contribute to a functional objective feels unnecessary and unwarranted \cite{clark_what_2019,cowan2017can,doyle_mapping_2019}, even unnatural for the given context. We should not overreact to this evidence and throw the champagne out with the cork however. Naturalness without functionality feels inappropriate, yes, but this alone does not mean that naturalness is inappropriate. Instead, we propose a different approach for setting appropriate expectations. We believe that naturalness and appropriateness can co-exist if we allow CUIs to embrace their incompetence! 

Essentially, by combining naturalness and appropriateness as the emphasis behind design efforts, we argue for ecological design that: appreciates and reflects perceptions of CUIs as a distinct type of dialogue partner with idiosyncratic communicative tendencies; strives for congruence between a system’s component parts; is cognizant of the contexts systems are used in; and thus signals an appropriate conceptual model of system functionality and competence to users. We propose embracing incompetence as one way that this might be achieved.

\section{Embracing Incompetence}
It is clear that interactions with CUIs are not perfect. CUIs are prone to a variety of errors, ranging from misunderstanding what a person said, to misinterpreting the person's intent, to delivering inappropriate responses \cite{brennan_interaction_1995}. But this is not unique to HMD. Human dialogue partners can and frequently do experience just the same problem states as CUIs \cite{clark1987collaborating}. To commit conversational errors is not unnatural, but in the pursuit of ‘naturalness’ in CUI design conversational errors are largely overlooked. Indeed, when it comes to improving the conversational efficacy of systems, errors are treated as if they are to be avoided at all costs. This is evidenced in the rush among large technology companies to declare the lowest word error rates, with some proudly declaring figures as low a 6.3\% \cite{noauthor_microsoft_2016}; despite concerns about the environments these tests are conducted in and the lack of transparency in how figures are calculated \cite{szymanski-etal-2020-wer}. Designers have also been hesitant to introduce clear error messages that define why an error or miscommunication has occurred, instead opting for relatively ambiguous statements like Siri’s, “I’m sorry I don’t know that one”. Perhaps the intent is to trivialise or downplay system fallibility, but the consequence of this approach is poor discoverability and poor understanding of the system’s knowledge and processing state; issues that are never more acute in HMD than when miscommunication occurs.

Conversations are messy and conversational incompetence is a perfectly natural phenomenon. Appropriately setting expectations for it can be achieved through subtle reductions in speech synthesis quality that can in fact improve usability in certain contexts \cite{moore2017spoken}. Yet this high-realism type of naturalness in speech is not the sort of polished and confident communicators that CUI developers strive to deliver. Natural speech is riddled with markers of errors like disfluency \cite{redford_fluency_2015,corley_hesitation_2008} which listeners can pick up on and use to set appropriate expectations about their dialogue partner \cite{brennan_how_2001}]; whereas these cues are stripped away from most synthesized voices, leaving users to interpret an absence of incompetence markers as an absence of incompetence, which is in stark contrast to their interaction experiences. 

It is not inappropriate naturalness, we argue, that causes the gulf in expectations, but instead inappropriate, inconsistent and ill-defined applications of naturalness that are decidedly unnatural. People signal conversational incompetence – uncertainty or embarrassment about potential errors – through a variety of verbal and nonverbal cues in HHD. We package together embarrassment and uncertainty here as markers of conversational incompetence insofar as uncertainty in conversation is seen as an attitude in which a speaker is appraising the likelihood of error, whereas embarrassment is an emotional reaction to that appraisal \cite{fazio_attitudes_2007}. The aforementioned disfluencies, including hesitations, prolongations, and filled pauses like “um” and “uh” \cite{redford_fluency_2015} have been estimated as affecting approximately 6\% of words uttered in HHD \cite{fox_tree_effects_1995}. Some linguists even argue that these filled pauses ought to be considered meaningful words of their own right, as speakers use them to convey particular messages which listeners readily interpret \cite{clark_using_2002}. Among other communicative purposes, these fillers can communicate embarrassment in HHD, which can also be communicated with laughter filling the pause \cite{hawk_worth_2009}. Likewise, fillers and pauses are recognizable indicators of uncertainty in audio-only HHD, while smiles and funny faces can serve the same purpose as laughter when a listener can see the speaker’s face \cite{swerts_audiovisual_2005}. Indeed some research indicates that these sorts of facial gestures can be even stronger indicators of uncertainty than verbal cues like tone and word choice \cite{borras-comes_perceiving_2011}. Uncertainty can further be signalled by syntax as well, with the conditional mood (e.g. “I would like to say that …”), modal verbs (e.g. “might”), and tag-questions (e.g. “isn’t it?”) \cite{topka_situation_2013}. This multitude of cues we have in HHD for signalling our conversational incompetence, setting the appropriate expecation that error may occur, allows us to prepare the listener for errors directly in natural language, a communicative function largely absent in CUIs. 

Of course, communicative competence is also an important dimension of user perceptions in HMD \cite{branigan_role_2011, doyle_what_2021} and we are not suggesting a high error rate should just be accepted as 'good enough'. Instead we suggest a degree of incompetence can be signalled very subtly without significant disruption to the user achieving their goals. Some research on synthesized speech has already begun to demonstrate the viability of subtle makers of uncertainty in HMD. One study on perceptions of uncertainty with synthesized voices found that disfluencies like those in HHD can likewise cue human listeners to perceive a synthetic voice as sounding uncertain \cite{szekely_synthesising_2017}. Additionally, research on both English and German synthetic voices has found that prosodic qualities of the voices such as pronunciation and vocal effort level also have significant impact on human listeners’ and perceived uncertainty in those voices \cite{szekely_synthesising_2017,honemann_synthesizing_2016}. Other work in this area indicates that particular lexical cues of uncertainty, words like “well” and “right”, are more impactful than prosodic cues in signalling uncertainty with synthesized speech \cite{lai_what_2010}, but both types of cues together express uncertainty even more strongly \cite{lasarcyk_modelling_nodate}. It is worth noting that studies that introduce these markers of uncertainty into synthesized speech have not found differences in perceived naturalness between synthesized voices with or without these cues \cite{lasarcyk_modelling_nodate} nor between synthesized speech and human speech with the same markers \cite{wester_artificial_2015}. This evidence combines to deliver a key insight: CUIs can reliably communicate uncertainty, and doing so does not make the CUI or the interaction with it seem less natural to its dialogue partner.

We therefore see embracing incompetence - signalling embarrassment and/or uncertainty - as a promising avenue for setting appropriate expectations in conversation without entirely abandoning naturalness as a design goal. We merely suggest that natural features of human conversation are incorporated in a system and context sympathetic fashion: appropriate naturalness, or better again, an ecologically conscious approach. Indeed, the multitude of cues available to all sorts of CUIs - text based, voiced, and embodied - mean that CUIs, just like humans, can become masters of subtly expressing incompetence. Not just in a binary way, but in a multifaceted and subtly nuanced spectrum between confident certainty and humiliated bewilderment.

\section{Conclusion}
The CUI community has recognized a fundamental problem in how CUI users perceive CUIs resulting from a mismatch between the efficacy of their component parts. Calls for addressing appropriateness in the design of CUI communication styles have been well-founded, rooted in our understanding of how we perceive dialogue partners, and offer non-trivial usability benefits in certain contexts \cite{moore_appropriate_2017}. But recent suggestions about how to improve appropriateness which seek to move away from naturalness potentially misdiagnose the problem. They might also be circumvented by our tendency to anthropomorphize all manner of objects. It is not the case that CUIs have come to communicate too naturally, in a way that belies how unnatural they are in function. Instead, CUI design has failed to embrace their inherent incompetencies. Failure to give signals about uncertainty, embarrassment, or confusion has led users to interpret CUIs as signalling confidence, certainty, and competence. This highly unnatural mismatch contributes to significant usability issues due to the gap between user expectations and reality. We call for CUIs to embrace their incompetence, making their communication style simultaneously both more natural and more appropriate. In essence more ecologically valid. This approach, adding communicative signals rather than changing or removing signals, is available to CUIs of all sorts: text-based, voice-based, or embodied; task-oriented or open-domain; new prototypes or established commercial products. By embracing incompetence, CUIs add honest signals to communication, making that communication more appropriately aligned with the system’s capabilities and more like natural human-human dialogue rather than forcing developers to choose one aim or the other.

\begin{acks}
This research was conducted with the financial support of the ADAPT SFI Research Centre at University College Dublin. The ADAPT SFI Centre for Digital Content Technology is funded by Science Foundation Ireland through the SFI Research Centres Programme and is co-funded under the European Regional Development Fund (ERDF) through Grant \# 13/RC/2106\textunderscore{}P2.
\end{acks}

\bibliographystyle{ACM-Reference-Format}
\bibliography{Incompetence}

\appendix

\end{document}